\documentclass[twocolumn,showpacs,preprintnumbers,amsmath,amssymb,prl]{revtex4}

\usepackage{graphicx}
\usepackage{dcolumn}
\usepackage{bm}

\begin{document}


\title{Quantum simulation of manybody effects in steady-state nonequilibrium: electron-phonon coupled quantum dots}

\author{J. E. Han}
\affiliation{
Department of Physics, State University of New York at Buffalo, Buffalo, NY 14260, USA}

\date{\today}

\begin{abstract}
We develop a mapping of quantum steady-state nonequilibrium
to an effective equilibrium and solve the problem using a
quantum simulation technique. A systematic implementation
of the nonequilibrium boundary condition in steady-state
is made in the electronic transport on quantum dot structures.
This formulation of quantum manybody problem in nonequilibrium 
enables the use of existing numerical quantum manybody techniques.
The algorithm coherently demonstrates various transport behaviors
from phonon-dephasing to $I-V$ staircase and phonon-assisted 
tunneling.
\end{abstract}


\maketitle

Electronic transport in
nanoscale devices has recently attracted much attention for
its tremendous impact in modern electronics.
In nanoscale devices,
fundamental understanding of quantum physics
holds the key to a breakthrough in the field.
One of the central issues is to understand 
the increasing interplay of quantum manybody 
effect and the high-bias nonequilibrium.

Despite the intense research in the nanoscale transport, 
most of the theoretical tools are based on the Green
function technique~\cite{datta_book,wingreen,rammer,datta_prb},
and powerful numerical equilibrium techniques, which have
been essential in quantum manybody theories, have not been
useful due to the lack of nonequilibrium formulation. 
Application of such well-established numerical techniques, such
as quantum Monte Carlo (QMC) method~\cite{bss},
to nonequilibrium with full quantum manybody
effects will provide a breakthrough in the field.
The goal of this work is to formulate a general scheme of combining 
the nonequilibrium theory with numerical quantum manybody techniques
via a mapping of a nonequilibrium system to an effective equilibrium.
This method applied in electron-phonon coupled quantum dot
systems demonstrates various transport behavior such as
the phonon dephasing, phonon-assisted tunneling,  $I-V$ staircase 
in a single framework. The formulation shown is applicable
to other numerical techniques, such as nemerical renormalization 
group, exact diagonalization, etc.

\begin{figure}[bt]
\rotatebox{0}{\resizebox{!}{1.2in}{\includegraphics{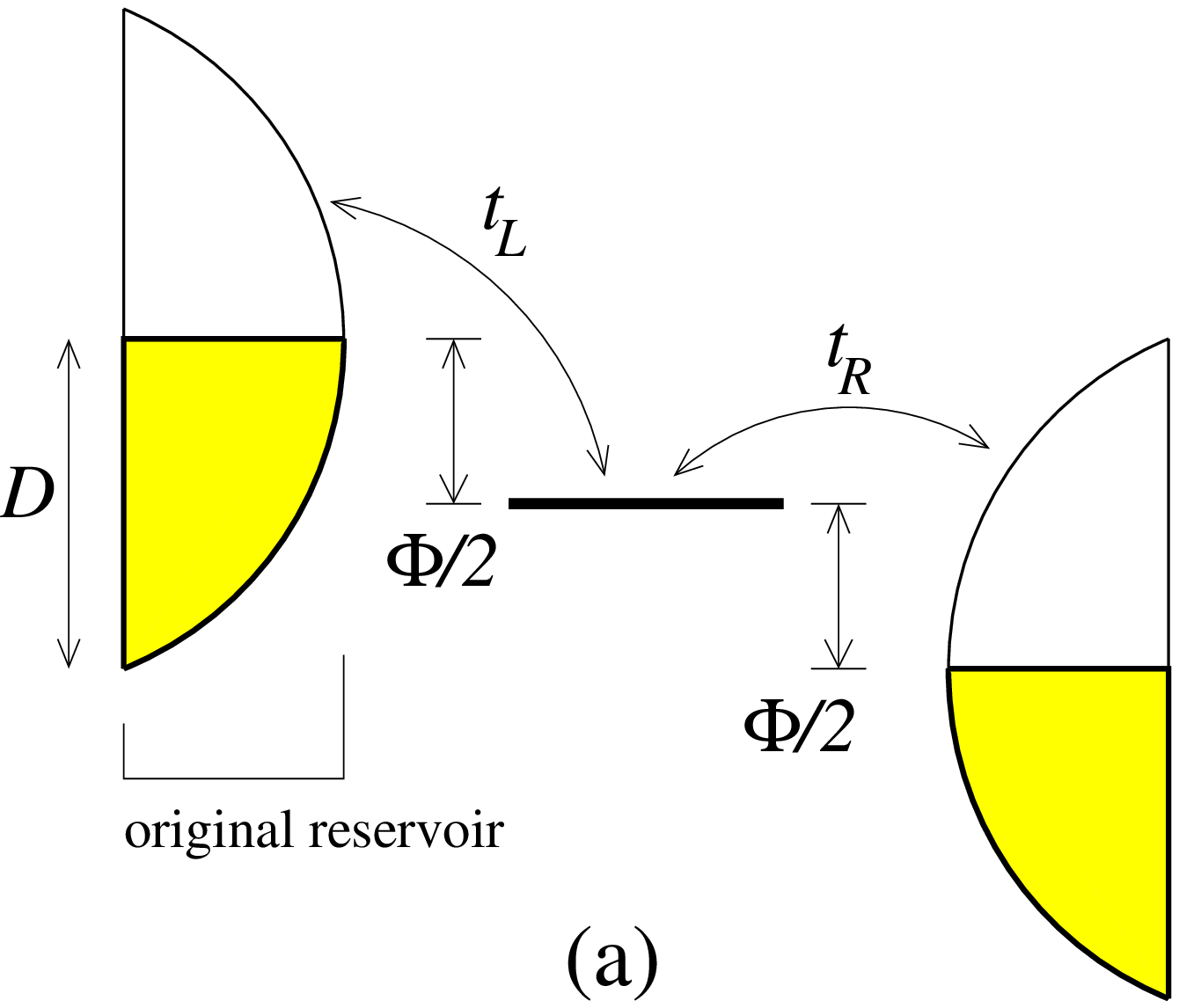}}}
\hfill
\rotatebox{0}{\resizebox{!}{1.2in}{\includegraphics{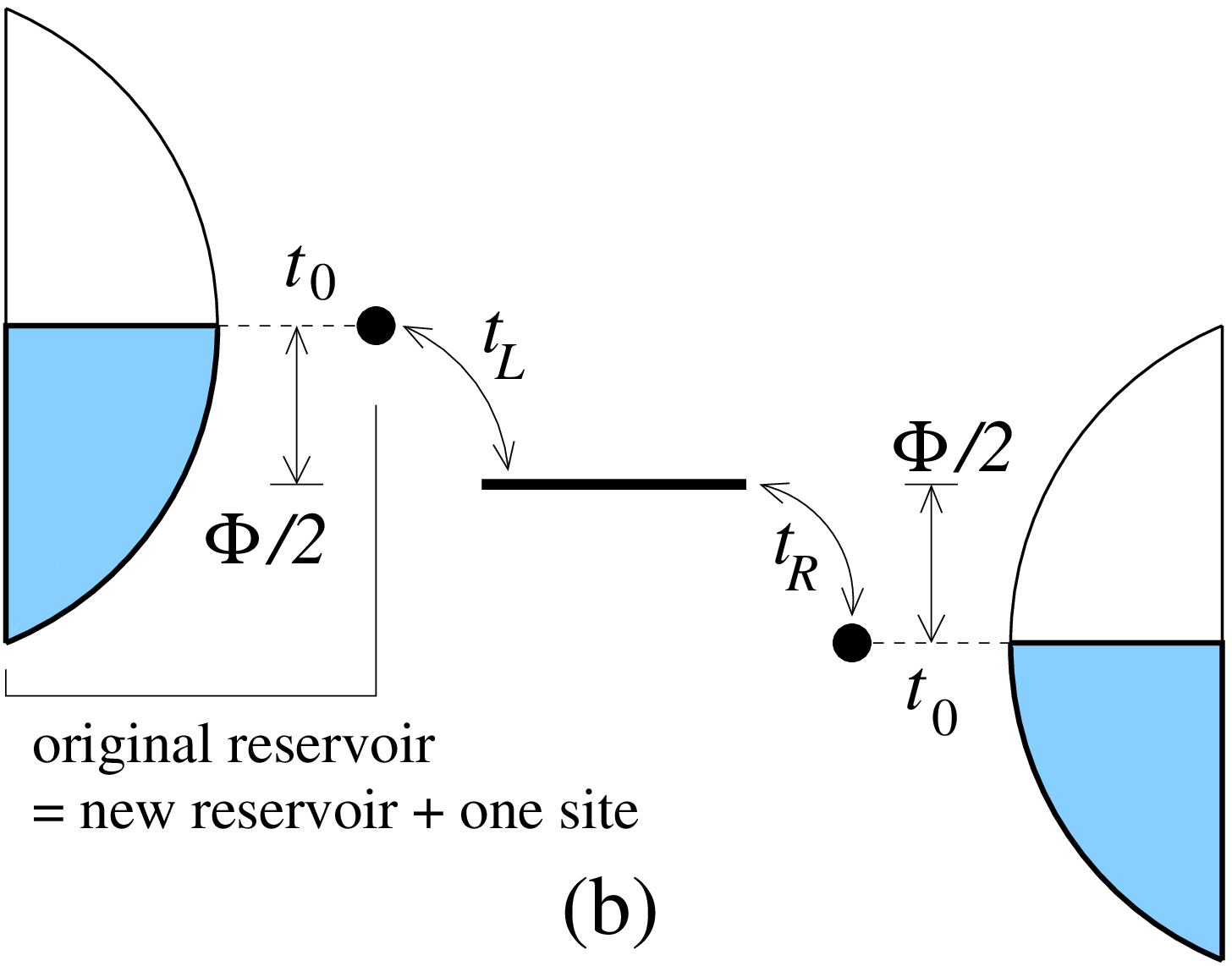}}}
\caption{
(a) Schematic diagram of a quantum transport device. The source 
and drain reservoirs are biased by a chemical potential
difference $\Phi$. The quantum dot (QD) level is set in the
middle for simplicity. 
(b) Each reservoir is mapped to a single site
and a new reservoir. The three discrete sites for
the QD and $L$- and $R$-sites are treated in fully quantum
mechanical manner.
}
\label{diagram}\end{figure}

Conventional diagrammatic theory is founded in adiabatic turn-on
of interaction. Here we begin from a very different point
that a steady-state nonequilibrium possesses the time-invariance,
where the transient states are damped out.
With this observation, Hershfield~\cite{hershfield}
has shown the existance of an effective Hamiltonian which
builds the nonequilibrium ensemble in steady-state.
The ensemble can be represented in terms of a modified 
partition function $Z_{\rm noneq}$ for nonequilibrium as
\begin{equation}
Z_{\rm noneq} = {\rm Tr}e^{-\beta(\hat{H}-\hat{Y})}
\quad\mbox{with}\quad\beta=1/T,
\end{equation}
where $\hat{H}$ is the usual Hamiltonian operator for the whole
system including the source (left, $L$), drain (right, $R$)
reservoirs and quantum dot (QD) (see FIG.~\ref{diagram}~(a)).
The additional operator $\hat{Y}$ accounts for the boundary
condition of chemical potential difference in the two reservoirs,
given by the bias $\Phi$. Once the bias operator
$\hat{Y}$ is constructed, one can apply the usual equilibrium
QMC technique~\cite{bss} to 
$\hat{H}_{\rm noneq}=\hat{H}-\hat{Y}$ 
as if $\hat{H}_{\rm noneq}$ describes an equilibrium. 
To obtain $I-V$ characteristics, for instance,
one simply needs to evaluate the expectation value of the
current operator $\hat{I}$,
\begin{equation}
I = \frac{{\rm Tr}\,\,\hat{I}e^{-\beta(\hat{H}-\hat{Y})}}{
  {\rm Tr}\,\,e^{-\beta(\hat{H}-\hat{Y})}}
 = \frac{ie}{h}\sum_{k\sigma}\frac{t_{Lk}}{\sqrt{\Omega}}
  \left[\langle c^\dagger_{Lk\sigma}d_\sigma\rangle
  -\langle d^\dagger_\sigma c_{Lk\sigma}\rangle\right],
\label{eq:I}
\end{equation}
where $c^\dagger_{Lk\sigma}$ ($c_{Lk\sigma}$) is the electron
creation (annihilation) operator of spin $\sigma$, 
the $k$-th continuum state 
in the $L$-reservoir, $d^\dagger_\sigma$
($d_\sigma$) the operator for QD state,
and $t_L$ the hopping integral (with the volume factor
$\Omega$) between the QD and the $L$-reservoir 
(see FIG.~\ref{diagram}(a)).

The central challenge is to construct the bias operator $\hat{Y}$.
Once a nonequilibrium ensemble is established via 
$e^{-\beta(\hat{H}-\hat{Y})}$,
the subsequent motion is governed by the time evolution operator 
$e^{-i\hat{H}t}$. Therefore the bias operator
$\hat{Y}$ must satisfy the commutation relation $[\hat{H},\hat{Y}]=0$.
Hershfield~\cite{hershfield} has shown that $\hat{Y}$ can be written 
in terms of the scattering state operator $\psi^\dagger_{\alpha k\sigma}$ 
($\alpha=L,R$) as~\cite{localized}
\begin{equation}
\hat{Y}=-\frac{\Phi}{2}\sum_{k\sigma}\left(
\psi^\dagger_{Lk\sigma}\psi_{Lk\sigma}-\psi^\dagger_{Rk\sigma}\psi_{Rk\sigma}
\right).
\label{eq:y}
\end{equation}
The scattering state creation operator $\psi^\dagger_{\alpha k\sigma}$
satisfies the Lippman-Schwinger
equation in the operator form,
\begin{equation}
  [\hat{H},\psi^\dagger_{\alpha k\sigma}]= \epsilon_{\alpha k}\psi^\dagger_{\alpha k\sigma} +i\eta(\psi^\dagger_{\alpha k\sigma}-c^\dagger_{\alpha k\sigma}),
\label{eq:psi}
\end{equation}
with $\epsilon_{\alpha k}$ the asymptotic continuum energy and $\eta$
an infinitesimal convergence factor.

There are two main issues to implement the above nonequilibrium ensemble.
First, to systematically construct $\hat{Y}$, 
the scattering state is expanded in terms of interaction as
$\psi^\dagger_{\alpha k\sigma}=\sum_n 
\psi^\dagger_{\alpha k\sigma,n}$~\cite{hershfield} satisfying
\begin{equation}
[\hat{V},\psi^\dagger_{\alpha k\sigma,n-1}]= (\epsilon_{\alpha k}+i\eta)
\psi^\dagger_{\alpha k\sigma,n}-[\hat{H}_0,\psi^\dagger_{\alpha k\sigma,n}],
\label{eq:expand}
\end{equation}
where the total Hamiltonian $\hat{H}=\hat{H}_0+\hat{V}$. $\hat{H}_0$
is the non-interacting and $\hat{V}$ the interacting part of the
Hamiltonian. 
The second and more serious issue is the non-locality of the operator
$\hat{Y}$. Since the one-particle eigenstate of the total Hamiltonian
is delocalized in space, the bi-products $\psi^\dagger_{\alpha k\sigma}
\psi_{\alpha k\sigma}$ in $\hat{Y}$
produce non-local terms even if the model Hamiltonian $\hat{H}$ is given
with local interactions. Despite the great potential of Hershfield's
paper~\cite{hershfield}, only limited developments~\cite{schiller,bokes} 
have been made so far due to such difficulties.

To make the following discussions concrete, 
let us consider an example: electron-phonon (el-ph)
coupled QD connected to two $L$, $R$-reservoirs. 
Spinless Hamiltonian for the system reads 
$\hat{H}=\hat{H}_0+\hat{V}=\hat{H}^{el}_0+\hat{H}^{ph}_0+\hat{V}$,
\begin{eqnarray}
 \hat{H}^{el}_0 & = & \sum_{\alpha k}\epsilon_{\alpha k}c^\dagger_{\alpha k}
c_{\alpha k} + \epsilon_d d^\dagger d
  +\sum_{\alpha k}\frac{t_{\alpha k}}{\sqrt{\Omega}}\left(
  d^\dagger c_{\alpha k} + h.c.\right) \nonumber \\
  \hat{H}^{ph}_0 & = & \frac{1}{2}\left(p^2+\omega_{ph}^2\varphi^2\right),
 \hat{V} = \alpha \varphi (d^\dagger d-\langle d^\dagger d\rangle),
\end{eqnarray}
where $\varphi$ is phonon amplitude, $p$ its conjugate
momentum, $\omega_{ph}$ phonon frequency, and 
$\alpha(\equiv g\sqrt{2\omega_{ph}})$ the el-ph coupling constant. 
As shown in FIG.~\ref{diagram}(a), we model the continuum states
to be shifted with the bias, i.e., $\epsilon_{Lk}=\epsilon_k+\Phi/2$ and
$\epsilon_{Rk}=\epsilon_k-\Phi/2$. We set
the particle-hole symmetric QD level ($\epsilon_d=0$) for simplicity.

Here we propose that we capture important manybody effects by
isolating the leading order non-local interactions as follows.
We make two observations in $\hat{H}-\hat{Y}$. 
First, the QD level is modulated
by phonons as $\epsilon_d(\varphi)=\epsilon_d+\alpha\varphi$.
The fluctuation of the QD level results in dephasing
of the current. Second, the hopping out of the QD to the
continuum is effectively modified as
$\epsilon_d(\varphi)$ is driven in and out of resonance with
respect to the reservoir chemical potentials.
As will be discussed later, fluctuations in the hopping integral 
result in phonon satellites in current.
With these observations, we isolate the 
first path of hopping from the QD by rewriting the each continuum
as a fictitious site and a new continuum, as depicted in 
FIG.~\ref{diagram}(b).
We then treat the coupling between the three sites fully quantum
mechanically and make mean-field approximations to terms
involving the continuum states. This procedure can be systematically
improved by inserting additional fictitious reservoir sites
to a better non-local approximation.

We have used a semi-circular density of state (DOS) for the
continua, $N(\epsilon_\alpha)
=2\sqrt{D^2-\epsilon_\alpha^2}/\pi D^2$, with the Fermi energy
$\epsilon_F=D$. The semi-circular DOS is particularly useful
since, with the intra-reservoir hopping $t_0=D/2$ (see FIG.~\ref{diagram}),
the resulting new DOS is identical
to the original DOS. In the non-interacting limit ($\hat{V}=0$),
the scattering state can be readily obtained by expanding 
$\psi^\dagger_{\alpha k}$ in terms of the basis states in 
Eq.~(\ref{eq:psi}) as
\begin{equation}
\psi^\dagger_{L k,0} = c^\dagger_{L k}
 +\frac{t_0}{\sqrt{\Omega}}
 \sum_\beta g^0_{L,\beta}(\epsilon_{Lk})c^\dagger_\beta,
\end{equation}
where $\beta=Lk', L, d, R, Rk'$ with $L,R$ denoting the 
fictitious reservoir sites, and $Lk', Rk'$ new continuum states.
$c^\dagger_d$ is the same as $d^\dagger$.
$g^0_{L\beta}(\epsilon)$ are
retarded Green functions propagating from state $L$ to $\beta$.
$\psi^\dagger_{R k,0}$ are similarly obtained.
From Eq.~(\ref{eq:I}), one can recover the Landauer-B\"uttiker
formula in the non-interacting limit.

We expand the scattering states up to harmonic el-ph coupling
via Eq.~(\ref{eq:expand}). With $[\hat{V},\psi^\dagger_{Lk,0}]
=(t_0/\sqrt\Omega)\alpha\varphi g^0_{Ld}(\epsilon_{Lk})d^\dagger$
and $\psi^\dagger_{\alpha k,1}$ expanded as
\begin{equation}
\psi^\dagger_{L k,1} = 
\varphi\sum_\beta a^{Lk}_\beta c^\dagger_\beta 
 +p\sum_\beta b^{Lk}_\beta c^\dagger_\beta,
\end{equation}
one obtains explicit expressions for $a_\beta, b_\beta$ after a 
straightforward calculation.
Up to the linear order of $(\varphi,p)$ we obtain 
$\hat{Y}=\hat{Y}_0+\hat{Y}_1$ with
$\hat{Y}_0=\Phi/2\sum_k(\psi^\dagger_{Lk0}\psi_{Lk0}-\psi^\dagger_{Rk0}
\psi_{Rk0})$, 
$\hat{Y}_1=\Phi/2\sum_k(\psi^\dagger_{Lk0}\psi_{Lk1}+\psi^\dagger_{Lk1}
\psi_{Lk0}-\psi^\dagger_{Rk0}\psi_{Rk1}-\psi^\dagger_{Rk1}\psi_{Rk0})$.
Now we make a mean-field approximations ($\varphi=0$)
on the terms which involve the continuum states 
$Lk'$, $Rk'$. Projected onto the discrete basis $(L,d,R)$, 
$\hat{Y}_1$ reads
\begin{equation}
\hat{Y}_1 = \varphi\hat{A}+p\hat{B}
      = \sum_{\beta,\gamma=L,d,R}\left(\varphi A_{\beta\gamma}
      +p B_{\beta\gamma}\right)c^\dagger_\beta c_\gamma,
\label{eq:AB}
\end{equation}
where the $3\times 3$ matrix $A$ is written as
\begin{eqnarray}
A_{\beta\gamma} & = & \frac{\Phi}{2}\sum_k\left[
 a^{Lk}_\beta \left(g^{0}_{L\gamma}(\epsilon_{Lk})\right)^*
+g^{0}_{L\beta}(\epsilon_{Lk})\left(a^{Lk}_\gamma\right)^*\right. \nonumber \\
 & & \left. -a^{Rk}_\beta \left(g^{0}_{R\gamma}(\epsilon_{Rk})\right)^*
-g^{0}_{R\beta}(\epsilon_{Rk})\left(a^{Rk}_\gamma\right)^*\right],
\end{eqnarray}
and $B$ is given by replacing $a^{\alpha k}_\beta$
by $b^{\alpha k}_\beta$.

We now sample the ensemble by treating
$\hat{H}_{\rm noneq}$ as in conventional QMC~\cite{bss}. 
We express the Boltzmann factor into the Trotter breakup~\cite{bss},
${\rm Tr}\,\exp(-\beta\hat{H}_{\rm noneq}) = {\rm Tr}[\exp(-\Delta\tau
\hat{H}_{\rm noneq})]^N$ with $\beta=N\Delta\tau$. The three discrete
sites ($L,d,R$) are considered as a generalized impurity and the Hirsch-Fye
algorithm~\cite{hirsch} is employed to integrate out the continuum 
states which are incorporated in the non-interacting
Green function $G^0_{\beta\gamma}(i\omega_n)$ at the Matsubara imaginary
frequency $\omega_n=(2n+1)\pi/\beta$ as
\begin{equation}
G^0_{\beta\gamma}(i\omega_n) = \!\!\!\sum_{\alpha=L,R; k}\!
\frac{g^0_{\alpha\beta}(\epsilon_{\alpha k})\left[g^0_{\alpha\gamma}
(\epsilon_{\alpha k})\right]^*}{i\omega_n-\epsilon_{\alpha k}+(-)^\alpha
\Phi/2}+
\sum_{m}\frac{c^m_\beta c^{m*}_\gamma}{i\omega_n-E_m},
\label{eq:green}
\end{equation}
where the possible localized states (with the
$m$-th energy $E_m$ and amplitudes 
$c^m_\beta$ for the $\beta$ state) are taken into account.
$(-)^L\equiv 1,\,(-)^R\equiv -1$. 
The localized states become important in the
narrow bandwidth limit.

The main difference from the usual equilibrium QMC is 
the $p$-term in Eq.~(\ref{eq:AB}). Integrating out the canonical
momentum $p$ inside time-slices of the Trotter-breakup
leads to~\cite{negele}
\begin{equation}
\langle \varphi_{\tau+\Delta\tau}|e^{-\Delta\tau(\frac12 p^2+p\hat{B})}
|\varphi_\tau\rangle\propto
e^{\Delta\tau
\left[-\frac12\left(\frac{\partial \varphi}{\partial\tau}\right)^2
+i\hat{B}_\tau\frac{\partial  \varphi}{\partial\tau}\right]},
\end{equation}
where $\hat{B}_\tau$ is the electronic operator 
at the imaginary time $\tau$.
Due to the second term $\hat{B}_\tau(\partial  \varphi /\partial\tau)$,
a temporally local update of the phonon field $\varphi_\tau$ results
in changes in $\hat{B}_{\tau-\Delta\tau},\hat{B}_{\tau},
\hat{B}_{\tau+\Delta\tau}$. 
The apparent broken time-reversal symmetry in the first-order time 
differentiation $(\partial  \varphi /\partial\tau)$
originates from the right-to-left flow of electrons induced by the
bias $\Phi$~\cite{feldman}.
The $(\partial\varphi/\partial\tau)$
term also shows bias-induced phonon fluctuations which become crucial
in phonon-assisted tunneling~\cite{tsui}.
$\Delta\tau=1$ in the following results.

\begin{figure}[bt]
\rotatebox{0}{\resizebox{3.0in}{!}{
\includegraphics{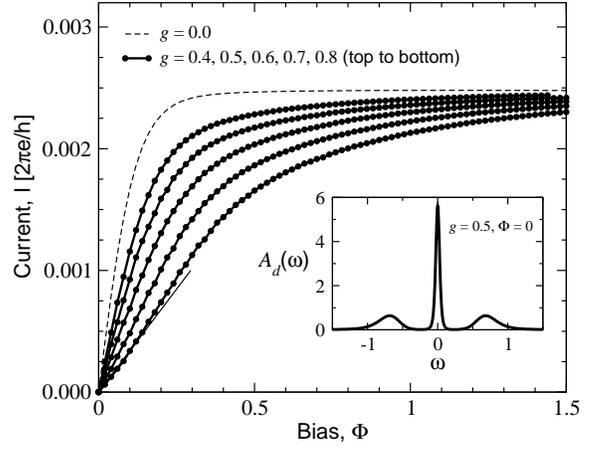}}}
\caption{Simulated $I-V$ characteristics in the broad band limit, 
$D\gg\omega_{ph}$. Parameters are $t_L=t_R=0.1,\,\omega_{ph}=0.5,\,
D=8.0$ and $T=1/32=0.03125$.
As the electron-phonon coupling constant $g$ is increased, the 
current decreased.  
Non-interacting phonon level aligns to the reservoir Fermi energy 
at bias $\Phi=2\omega_{ph} =1$. 
The phonon dephasing effect is so strong that
the phonon satellite peak is invisible at all coupling constant 
while the phonon satellites at zero bias (inset) are clearly
separated from the main peak.
}
\label{fig2}\end{figure}

$I-V$ curves in the broad band limit ($D\gg\omega_{ph}$)
are presented in FIG.~\ref{fig2}. The unit of energy is chosen
so that the bare phonon level ($\omega_{ph}=0.5$) aligns
with the Fermi energy of the $L$-reservoir at unit bias ($\Phi=1$).
The current monotonically increased with $\Phi$
at all el-ph coupling constant $g$.
The current converges toward the well-known 
large-band, large-bias limit
(dashed line for $g=0$)~\cite{spin}
\begin{equation}
I=\frac{2e}{\hbar}\frac{\Gamma_L\Gamma_R}{\Gamma_L+\Gamma_R},
\quad\mbox{with\ } \Gamma_a=\pi t_\alpha^2 N_\alpha(0).
\end{equation}

At zero bias,
the QD spectral function $A_d(\omega)$ with real frequency $\omega$
(see inset) by using the maximum 
entropy method~\cite{mem} shows that the phonon satellite
peaks are clearly separated from the main peak~\cite{analytic_cont}. 
From this, one 
might have expected that there would be a double-step 
$I-V$ characteristics, while the computed curves do not show 
such distinct features.
This suggests that the actual decay
rate of the phonon satellite under finite bias
($\Phi\approx 2\omega_{ph}$) is much larger than the zero bias case because
the phonon level is now close to a resonance with the reservoir 
Fermi energy while at zero bias finite particle-hole excitation 
energies are required for phonons to decay.

However, upon close inspection of the curves, 
one finds that the line shapes change slightly. 
To guide the eye, a tangential line is drawn for $g=0.8$ from the 
zero bias where the renormalized coherent current dominates.
As $\Phi$ increases, it quickly gives way to 
the transport which involves incoherent phonon excitations. 
In the broad band limit, the dominant role of phonons
is the dephasing effect from the fluctuation of the QD level 
$\epsilon_d(\varphi)=\epsilon_d+\alpha\varphi$ via the
$\hat{Y}_0$ term, instead of 
phonon excitations serving as well-defined discrete levels.
Here the current is rather insensitive to the choice of 
$\hat{Y}_1$.

\begin{figure}[bt]
\rotatebox{0}{\resizebox{3.0in}{!}{
\includegraphics{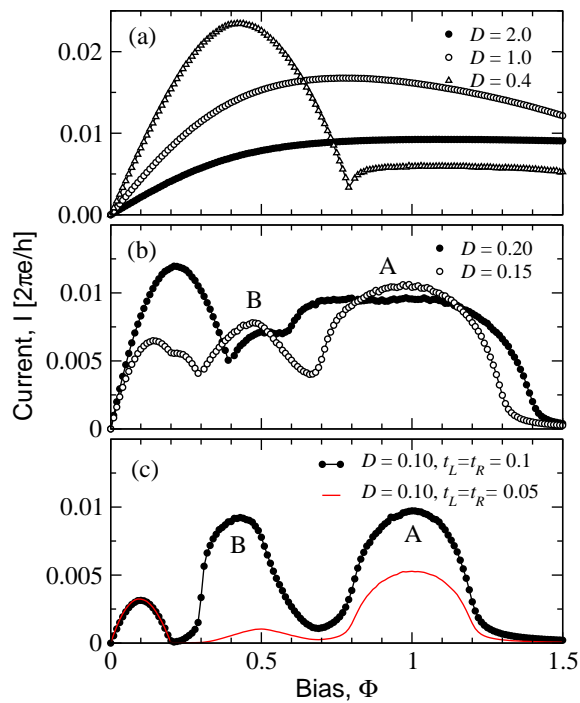}}}
\caption{$I-V$ characteristics for narrow bandwidth limit.
$t_L=t_R=0.1$ (except for (c)), $\omega_{ph}=1$, $g=0.2$ and 
$T=1/12=0.0833$.
(a) The saturated current in the large bandwidth limit
gradually turns into negative differential resistance regime
as $\Phi$ becomes comparable to $D$.
(b) Three-peak structure emerges as $D$ gets further reduced.
The peaks correspond to ballistic coherent transport, 
phonon-assisted tunneling (B, $\Phi=\omega_{ph}$)
and sequential off-resonant transport (A, $\Phi=2\omega_{ph}$).
(c) Reduction of tunneling parameter $t_L,t_R$ dramatically suppresses 
the phonon-assisted tunneling compared to the sequential current.
}
\label{fig3}\end{figure}

In the narrow band limit, one obtains a negative differential
resistance (NDR) behavior~\cite{tsui}. 
FIG.~\ref{fig3} shows the results when the
bandwidth $D$ is gradually decreased. Parameters are $\omega_{ph}=0.5,\,
t_L=t_R=0.1$ (except in (c) as shown) and $T=1/12=0.0833$.
As $D$ becomes comparable to $\Phi$, a gradual NDR shows 
up due to the decreasing overlap of $L-R$ DOS. 
As $D$ is further decreased a distinct discrete three-peak feature emerges 
as shown in FIG.~\ref{fig3}(b-c). 
The peak at near the zero bias is the coherent current corresponding to
the ballistic transport renormalized by the interaction. 
As $D$ is reduced well below $T$, the maximum height of the
ballistic current gets reduced due to the thermal dephasing.

The phonon satellite peak at higher bias ($\Phi=2\omega_{ph}=1$,
marked as A in the plot) becomes evident in FIG.~\ref{fig3}(b). 
The electronic
transport in this regime is sequential~\cite{datta_book}, 
i.e., electron from the 
$L$-reservoir tunnels to the QD by emitting a phonon and loses 
its phase information before it tunnels out to the $R$-reservoir
after emitting another phonon.
The width of the structure is twice as large as the full bandwidth,
$4D$, as expected. It is interesting that the phonon structures
for $D=0.20$ have the familiar stair-case $I-V$ curve, which shows
that the phonon satellite states act like discrete current channels
due to the reduced decay paths to continuum,
in contrast to the large bandwidth limit where the phonon acts
more as a dephaser than as a discrete level. We emphasize that the
phonon satellite features $A$ and $B$ are due to 
the first order correction to $\hat{Y}$, particularly the
$p$-term in $\hat{Y}_1$.

\begin{figure}[bt]
\rotatebox{0}{\resizebox{3.0in}{!}{
\includegraphics{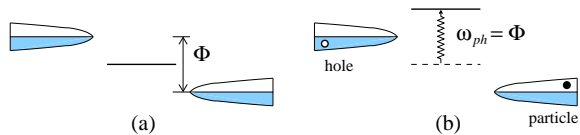}}}
\caption{(a) Initial and (b) final states in the phonon assisted
tunneling process. The particle-hole and phonon excitations are 
resonant at the bias $\Phi=\omega_{ph}$.
}
\label{fig4}\end{figure}

An intriguing feature in the $I-V$ curves is the phonon-assisted
tunneling marked as $B$ in FIG.~\ref{fig3}. As depicted in FIG.~\ref{fig4},
the tunneling happens by energy-exchange between a particle-hole
pair across the device and a phonon on the QD at the bias 
$\Phi=\omega_{ph}$. The curves at $D=0.4, 0.2$ resemble the experimental
results~\cite{tsui}. This is a direct consequence of
the coherence built in the $\hat{Y}$ operator, Eq.~(\ref{eq:AB}), and
should be distinguished from the sequential tunneling. To demonstrate
the point, a calculation with reduced tunneling parameters $t_L=t_R=0.05$
is shown as a thin line in FIG.~\ref{fig3}(c). Since the phonon-assisted
tunneling is through a coherent electron-hole state,
the tunneling amplitude is proportional to its cross-lead overlap
$t_L t_R$,
while the amplitude of the sequential tunneling is proportional
to $t_L$ or $t_R$ separately due to the uncorrelated sequential
tunneling events. The much reduced peak $B$ supports the argument.

We have formulated a quantum simulation algorithm for steady-state
nonequilibrium and
have shown that the method reproduced the most important
physics in the electron-phonon coupled quantum dot system.
In combination with other numerical manybody techniques, the
formulation of nonequilibrium shown here will provide
a critical step toward a more complete theory of steady-state
nonequilibrium.

\begin{acknowledgments}
I thank helpful discussions with A. Schiller, F. Anders
and P. Bokes.
I acknowledge support from the National Science 
Foundation DMR-0426826 and computational resources from the Center
for Computational Research at SUNY Buffalo.
\end{acknowledgments}


\begin{thebibliography}{*}
\bibitem{localized} In the narrow band limit, there are localized states
which cannot be included in $\psi_{\alpha k\sigma}$. However they do not
contribute to $\hat{Y}$.
\bibitem{spin} This expression does not include summation over spins.
\bibitem{analytic_cont}
Direct analytic continuation at finite bias does not
have an obvious physical meaning since the Green functions 
(see Eq.~(\ref{eq:green})) have poles from $\hat{H}-\hat{Y}$, 
not $\hat{H}$.
\end{thebibliography}
\end{document}